 \documentclass[pmlr,twocolumn,10pt]{jmlr} 

\usepackage{booktabs}
\usepackage{siunitx}

\usepackage[switch]{lineno}
\usepackage{enumitem}

\newcommand{\equal}[1]{{\hypersetup{linkcolor=black}\thanks{#1}}}

\theorembodyfont{\upshape}
\theoremheaderfont{\scshape}
\theorempostheader{:}
\theoremsep{\newline}

\jmlrvolume{297}
\jmlryear{2025}
\jmlrworkshop{Machine Learning for Health (ML4H) 2025} 

  \title[Prostate-VarBench: A Benchmark with Interpretable TabNet Framework for Prostate Cancer]{Prostate-VarBench: A Benchmark with Interpretable TabNet Framework for Prostate Cancer Variant Classification}

 \author{%
  \Name{Abraham Francisco Arellano Tavara}\equal{These authors contributed equally} \Email{aa107@illinois.edu}\\
  \Name{Umesh Kumar}\footnotemark[1]  \Email{umesh2@illinois.edu}\\
  \Name{Jathurshan Pradeepkumar} \Email{jp65@illinois.edu}\\
  \Name{Jimeng Sun} \Email{jimeng@illinois.edu}\\
  \addr University of Illinois Urbana-Champaign, IL, USA
 }

\begin{document}

\maketitle

\begin{abstract}

Variants of Uncertain Significance (VUS) limit the clinical utility of prostate cancer genomics by delaying diagnosis and therapy when evidence for pathogenicity or benignity is incomplete. Progress is further limited by inconsistent annotations across sources and the absence of a prostate-specific benchmark for fair comparison. We introduce Prostate-VarBench, a curated pipeline for creating prostate-specific benchmarks that integrates COSMIC (somatic cancer mutations), ClinVar (expert-curated clinical variants), and TCGA-PRAD (prostate tumor genomics from The Cancer Genome Atlas) into a harmonized dataset of 193,278 variants supporting patient- or gene-aware splits to prevent data leakage. To ensure data integrity, we corrected a Variant Effect Predictor (VEP) issue that merged multiple transcript records, introducing ambiguity in clinical significance fields. We then standardized 56 interpretable features across eight clinically relevant tiers, including population frequency, variant type, and clinical context. AlphaMissense pathogenicity scores were incorporated to enhance missense variant classification and reduce VUS uncertainty. Building on this resource, we trained an interpretable TabNet model to classify variant pathogenicity, whose step-wise sparse masks provide per-case rationales consistent with molecular tumor board review practices. On the held-out test set, the model achieved $89.9\%$ accuracy with balanced class metrics and the VEP correction yields an $6.5\%$ absolute reduction in VUS.

\end{abstract}

\paragraph*{Data and Code Availability}
We integrate variants from three publicly available databases: COSMIC Mutant Census v102 (GRCh38 assembly)~\citep{10.1093/nar/gky1015}, ClinVar (accessed through Ensembl VEP)~\citep{McLaren2016TheEnsemblVariantEff}, and TCGA-PRAD~\citep{TCGA2015_PRAD_Cell} from The Cancer Genome Atlas \url{https://www.cancer.gov/tcga}. All source databases are publicly accessible to researchers. Our complete implementation, including data processing pipelines, TabNet training code, and attention analysis scripts, is available through our GitHub repository: \url{https://github.com/AbrahamArellano/uiuc-cancer-research/}. The repository includes comprehensive documentation, reproducible workflows, and example datasets to enable community adoption and validation.

\paragraph*{Institutional Review Board (IRB)}
This research uses exclusively publicly available genomic databases (COSMIC, ClinVar, TCGA-PRAD) containing de-identified variant information. No human subjects research or institutional approval was required for this computational analysis of existing public datasets.

\section{Introduction}
\label{sec:intro}

Prostate cancer is the most commonly diagnosed malignancy in men, with recent projections estimating approximately $300K$ new cases in 2025, accounting for nearly $30\%$ of male cancer diagnoses in the United States~\citep{siegel2025cancer}. Advances in genomic profiling of prostate tumors have enhanced our understanding of the disease and play a crucial role in guiding therapeutic decisions~\citep{hatano2021genomic}, including the use of PARP inhibitors~\cite{xia2021role}, androgen receptor-targeted therapies~\citep{crawford2018androgen}, and immunotherapy~\citep{fay2020immunotherapy}. However, the clinical utility of current assays is hindered by the high prevalence of \emph{Variants of Uncertain Significance (VUS)}~\citep{mellgard2024variants}, genomic alterations for which there is insufficient or conflicting evidence to classify them as either benign or pathogenic~\citep{richards2015standards}. Prostate cancer genomic testing frequently yields VUS in $30-50\%$ of cases~\citep{nicolosi2019prevalence}. Clinicians frequently encounter inconclusive reports that require time-intensive expert curation and, in many cases, repeat testing—consequences that contribute to delays in care and disparities in access to targeted therapies.

Deep learning has demonstrated strong performance across a range of predictive modeling tasks in oncology, from cancer diagnosis~\citep{kumar2024automating,tandon2024systematic}, prognosis~\citep{zhu2020application}, and treatment recommendations~\citep{xia2024deep}. It offers a promising path toward more consistent and scalable variant classification. However, clinical adoption of such models hinges not only on predictive accuracy but also on inherent interpretability. While black‑box models (e.g., deep neural networks, deep ensembles) may achieve strong predictive performance, their reliance on post‑hoc explanation techniques can decouple model reasoning from clinical rationale and complicate regulatory review~\citep{Rudin2019NatMI,FDA2021GMLP,Adebayo2018Sanity}. Despite being neural-based models, methods such as TabNet~\citep{arik2020tabnetattentiveinterpretabletabular} provide sequential, sparse feature-selection produced by its attention mechanism, enabling case-level rationales that align with how molecular tumor boards review evidence.

However, progress in modeling prostate cancer is limited by the lack of a prostate-specific benchmark and evaluation protocol for variant classification. Real-world resources are fragmented and inconsistently annotated; even widely used databases, such as ClinVar~\citep{landrum2016clinvar}, can contain conflicting classifications for the same variant across different submitters, complicating ground-truth construction and reproducibility~\citep{so2024reinterpretation}. Moreover, data leakage is a documented widespread problem in ML-based science, affecting at least 294 papers across 17 fields~\citep{kapoor2023leakage}, with variant prediction specifically hindered by multiple types of circularity~\citep{grimm2015evaluation}. Many genomic ML studies risk information leakage without patient- or gene-aware splits and rarely include temporal validation~\citep{FDA2025_AIEnabledDSF_Draft,FDA2021_GMLP}.




To address the above challenges, we propose Prostate-VarBench, a curated prostate-specific benchmark creation pipeline and an interpretable deep-learning framework for prostate cancer variant classification. Our contributions are as follows: 
\begin{itemize}[left=0pt]
    \item \textbf{Prostate-VarBench :} We curated a prostate-specific, leakage-controlled benchmark creation pipeline to support reproducible evaluation by integrating large-scale resources, including COSMIC~\citep{10.1093/nar/gky1015}, ClinVar~\citep{landrum2016clinvar}, and TCGA-PRAD~\citep{TCGA2015_PRAD_Cell}.
    \item \textbf{Interpretable prostate-specific classifier :} We present the first TabNet framework tailored to prostate cancer variant classification with native, step-wise attention masks that provide case-level rationales suitable for clinical review.
    \item \textbf{Systematic data-quality correction:} We introduce a Variant Effect Predictor (VEP) annotation concatenation correction to address a systematic data quality issue that affects clinical significance fields and downstream tasks. 
    \item \textbf{Tiered clinical feature ontology :} A compact, 8-tier schema (VEP-corrected, core VEP, AlphaMissense, population genetics, functional predictions, clinical context, variant properties, prostate biology) that supports tier-level and feature-level interpretation.
    \item \textbf{Empirical performance, impact, and deployability :} After leakage removal, the model attains $\sim89.9\%$ test accuracy with balanced class metrics, and its attention masks highlight clinically meaningful drivers (e.g., \textit{VAR\_SYNONYMS}, AlphaMissense, clinical context). We also observe a $\sim6.5\%$ reduction in VUS and report efficient inference suitable for clinical workflows, with all code and analyses packaged for reproducible evaluation.
\end{itemize}

\section{Related Work}


\subsection{Traditional genomic variant classification—limitations.}
Classical predictors (SIFT, PolyPhen\mbox{-}2, CADD) and meta\mbox{-}predictors (MetaSVM/MetaLR) remain widely used but emphasize conservation/sequence features, show ancestry-related biases, and yield task-agnostic scores that lack therapeutic context~\citep{Ng2003SIFT,Adzhubei2010PolyPhen2,Kircher2014CADD,Dong2015MetaSVM,Popejoy2016GenomicsDiversity}.
In clinical pipelines, VUS remain common—\(\sim41\%\) of tested individuals carry \(\ge 1\) VUS and \(31.7\%\) receive VUS-only results; complicating decisions even under ACMG/AMP standards~\citep{Chen2023JAMANetOpenVUS,Richards2015ACMGAMP}.
These gaps motivate disease-specific, \emph{inherently interpretable} models tailored to prostate cancer genomics.

\subsection{Machine Learning Evolution: Performance vs. Interpretability Trade-offs}  
Recent machine learning approaches demonstrate competitive performance—XGBoost implementations~\citep{Khandakji2022BRCA2XGB} achieve ROC-AUC of 0.93 in prostate cancer variant prediction, while deep CNNs achieve F1-scores exceeding 0.88 across cancer types. However, these gains create critical clinical deployment barriers: (1) \emph{post-hoc explanation dependency}—requiring SHAP or LIME methods that may not reflect actual model reasoning, compromising regulatory approval, (2) \emph{black-box opacity}—preventing molecular tumor board integration where clinicians require transparent rationales~\citep{Tamborero2022_MTBP}, and (3) \emph{generalization challenges}—pan-cancer models lack prostate-specific optimization for androgen receptor pathway analysis and DNA repair deficiency assessment critical for PARP inhibitor selection. Regulatory frameworks demand explainable AI for medical devices~\citep{FDA2025_AIEnabledDSF_Draft,FDA2021_GMLP}, creating an interpretability-performance gap that limits clinical adoption. TabNet~\citep{arik2020tabnetattentiveinterpretabletabular} addresses this trade-off through sequential attention mechanisms providing inherent explainability while maintaining competitive performance.

\subsection{Critical Research Gap: Interpretable AI for Prostate Cancer Genomics}
While interpretable architectures such as TabNet show promise for genomic applications, we found no prior work applying TabNet specifically to \emph{prostate} cancer variant classification with a clinically grounded feature hierarchy—leaving a clear, disease-specific gap. Current prostate cancer AI efforts concentrate on imaging and pathology, whereas genomics pipelines remain dominated by non-interpretable or post-hoc–explained models. In a pan-cancer, tumor-only WES benchmark spanning seven TCGA cohorts (105 patients), TabNet performed competitively while preserving native mask-based explanations, supporting its suitability for tabular genomics tasks that require built-in interpretability \citep{McLaughlin2023NPJ,Arik2021TabNet}.

\section{Methods}
\begin{figure*}[t!]
\floatconts
    {fig:tabnet-method}
    {\caption{Experimental setup and TabNet architecture overview. Input sources and leakage controls feed tiered features (VEP-corrected, core VEP, AlphaMissense, population, functional, clinical context, variant properties, prostate biology). TabNet executes six decision steps with sparse feature masks; step logits are aggregated into class probabilities. Masks are retained for per-variant explanations and tier-level analyses referenced in Section~\ref{subsec:attention-analysis}}}
    {\centering \includegraphics[width=0.9\linewidth]{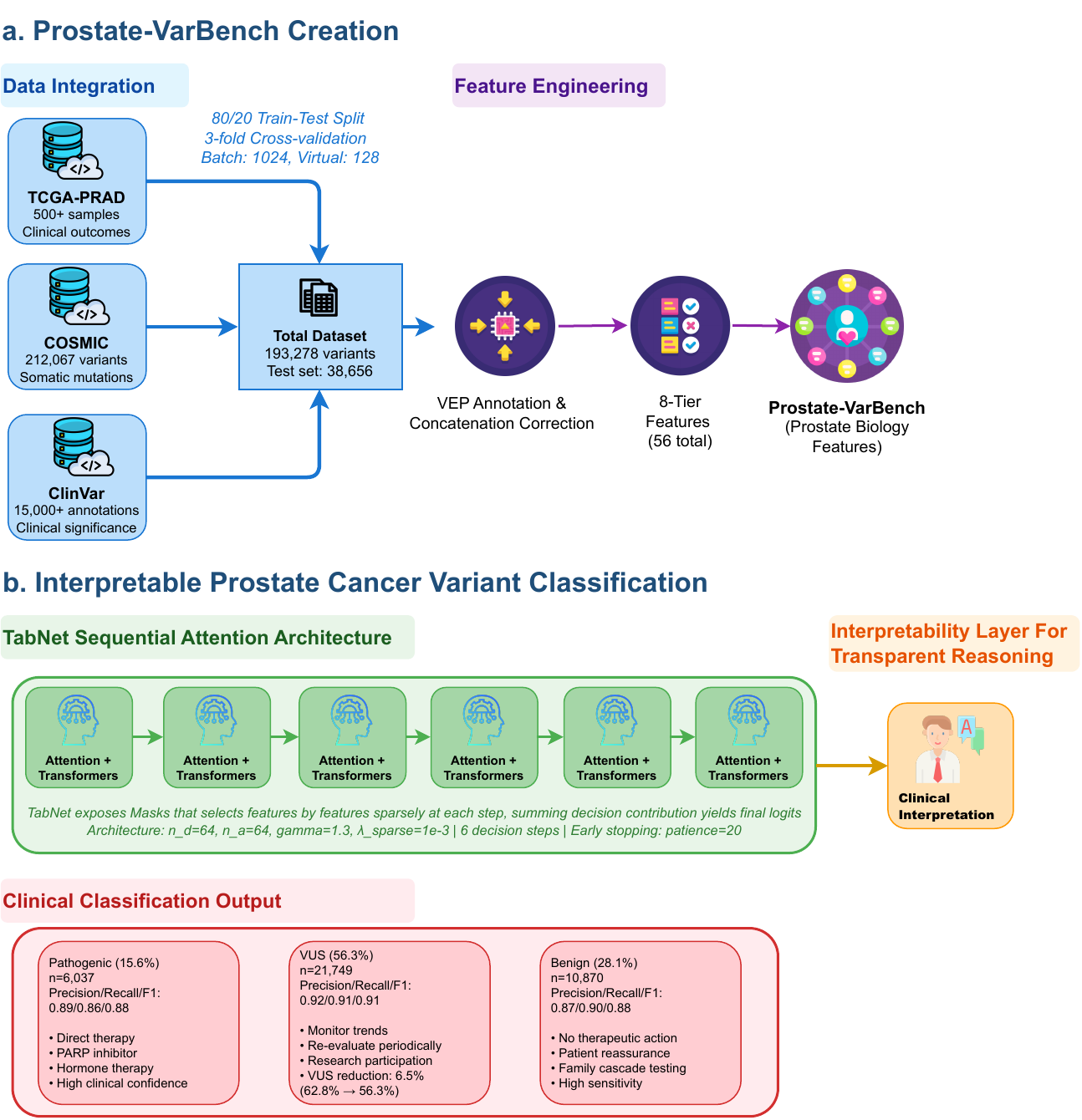}
    \vspace{-0.6cm}}
    \vspace{-0.8cm}
\end{figure*}

\begin{table*}
\centering
\small
\caption{Feature Engineering and Clinical Hierarchy. Attention shares are tier-level importances from TabNet. For more details, see Section~\ref{subsec:attention-analysis}.}
\resizebox{\linewidth}{!}{%
\begin{tabular}{l l r l l r l}
\toprule
Tier & Tier name & \# & Representative fields & Attn.\% & Clinical rationale \\
\midrule
T1 & VEP-Corrected & 4 & Consequence; DOMAINS; VAR\_SYNONYMS & 29.8 & Post-correction IDs/evidence \\
T2 & Core VEP & 10 & SYMBOL; BIOTYPE; HGVSp & 11.0 & Gene/transcript context \\
T3 & AlphaMissense & 2 & amissense\_score; amissense\_class & 14.0 & AI missense pathogenicity \\
T4 & Population Gen. & 17 & AF; AFR\_AF; EUR\_AF & 9.0 & Ancestry frequencies \\
T5 & Functional Pred. & 6 & IMPACT; SIFT; PolyPhen & 7.8 & In-silico effects \\
T6 & Clinical Context & 5 & SOMATIC; PHENO; EXON/INTRON & 14.8 & Setting/locus context \\
T7 & Variant Properties & 8 & is\_snv; is\_lof; variant\_size & 11.1 & Structural attributes \\
T8 & Prostate Biology & 4 & important\_gene; DNA/MMR/hormone & 2.5 & Prostate pathways \\
\bottomrule
\end{tabular}
}

\label{table:tiers}
\end{table*}

\subsection{Prostate-VarBench: Benchmark Construction}
\label{subsec:prostate-varbench}
In this section, we provide details on the construction of \textit{Prostate-VarBench}, which is a comprehensive data pipeline designed to transform heterogeneous genomic databases into a standardized, interpretable machine learning benchmark for prostate cancer variant classification (~\figureref{fig:tabnet-method}a). Our methodology follows a four-stage progression: (1) multi-source data integration with prostate-specific filtering, (2) systematic annotation enhancement and quality correction, (3) clinically-informed feature engineering across hierarchical tiers, and (4) rigorous quality control with leakage prevention to ensure realistic performance assessment. This pipeline addresses critical data quality issues while establishing reproducible evaluation standards for precision oncology applications.

\paragraph{Multi-Source Data Integration.}
\label{subsubsec:data-integration-annotation}
The foundation of our framework integrates three complementary genomic databases, each contributing distinct perspectives on prostate cancer variants. From COSMIC Mutant Census v102 (GRCh38 assembly)~\citep{10.1093/nar/gky1015}, we extracted somatic mutations using prostate-specific filtering criteria: direct tissue annotation matching, prostate cancer cell line identification (PC3, LNCaP, DU145, 22Rv1, VCaP, LAPC4, C4-2, MDA-PCa series), and TCGA-PRAD sample recognition, yielding 212,067 prostate-relevant variants across $10,408$ genes. ClinVar's GRCh38 VCF release~\citep{10.1093/nar/gkv1222, landrum2016clinvar} contributed 191,891 clinically annotated variants with established pathogenicity classifications, while TCGA-PRAD~\citep{TCGA2015_PRAD_Cell} provided $20,054$ mutations from 500+ patients with real-world clinical context.


\paragraph{Systematic VEP Annotation and Correction Pipeline.}
Building on the integrated dataset, comprehensive annotation employed Ensembl VEP v110.0~\citep{McLaren2016TheEnsemblVariantEff}, yielding 88 fields. Our methodological innovation identified and corrected systematic VEP \emph{concatenation} artifacts affecting 15.8\% of clinical significance and other multi-transcript fields. The implementation detects ampersand-concatenated values, reconstructs per-transcript records, and selects a single representative via a fixed clinical priority (MANE/canonical transcript $\rightarrow$ highest \texttt{IMPACT}/\texttt{Consequence} severity $\rightarrow$ presence of known variant IDs such as rsID/COSMIC/ClinVar); residual evidence is summarized as counts/flags (e.g., number of \texttt{DOMAINS}, distinct \texttt{VAR\_SYNONYMS}). Corrections are logged with pre/post values, validated for schema consistency, and applied \emph{prior} to feature engineering and target curation. All variants were then harmonized to unique keys \((\text{chr},\text{pos},\text{ref},\text{alt})\) with canonical transcript mapping, deduplicated, and consolidated into the final 193{,}278-variant corpus.

\paragraph{Clinical Feature Engineering and Hierarchical Design.}
\label{subsubsec:feature-engineering}
Following annotation quality assurance, we engineered 56 features across 8 clinical tiers designed (see Table~\ref{table:tiers}) to mirror established variant interpretation workflows while maintaining computational tractability. This hierarchical approach captures comprehensive genomic context from molecular mechanisms to clinical actionability, enabling both feature-level and tier-level interpretability analysis.

Each feature tier addresses specific clinical reasoning components: VEP-corrected features provide post-correction evidence strength, while AlphaMissense integration represents state-of-the-art AI pathogenicity assessment~\citep{cheng2023alphamissense}. Population genetics tiers enable ancestry-aware interpretation, and prostate biology features capture cancer-specific pathway disruptions essential for therapeutic selection. Categorical fields underwent label encoding for TabNet embedding, while numeric fields received standardization preprocessing.

\paragraph{Quality Control and Target Creation.}
\label{subsubsec:quality-control}
With engineered features established, rigorous quality control addressed data leakage—a widespread problem affecting hundreds of ML studies~\citep{kapoor2023leakage, grimm2015evaluation}. Initial model development revealed circular logic where clinical significance (\texttt{CLIN\_SIG}) appeared both as input feature and target component; a form of direct target leakage~\citep{kapoor2023leakage}—artificially inflating accuracy to 98\%. We eliminated this leakage by excluding all target-informative fields from the feature set, reducing from 57 to 56 features while implementing hierarchical target creation.

Our multiclass targets \(\{\text{Benign},\ \text{VUS},\ \text{Pathogenic}\}\) follow clinical reasoning hierarchy: primary classification from \texttt{CLIN\_SIG} when available, fallback to AlphaMissense class predictions, and final fallback to VEP \texttt{IMPACT} severity rules. This approach ensures realistic clinical scenarios where multiple evidence sources inform variant interpretation. The resulting performance drop to 89.9\% confirmed successful leakage elimination while maintaining clinically achievable accuracy benchmarks.

\subsection{Evaluation Framework}
\label{subsubsec:benchmark-construction}
The final pipeline stage establishes \textit{Prostate-VarBench} as a standardized evaluation framework enabling reproducible research across institutions. Data splitting employed stratified sampling with fixed random seeds, ensuring no variant key overlap across train/validation/test partitions to prevent Type 2 circularity~\citep{grimm2015evaluation} while maintaining class balance. This approach prevents both data leakage and optimistic performance estimates that plague genomic ML studies.

Our evaluation framework accommodates diverse baselines, including Logistic Regression (LR), XGBoost (XGB), Random Forest (RF), Support Vector Machine (SVM), and Multi-Layer Perceptron (MLP), through shared preprocessing pipelines, enabling a fair comparison against TabNet's interpretable architecture. Comprehensive metrics include Balanced Accuracy for class-imbalanced scenarios, Cohen's \(\kappa\) for inter-rater agreement simulation, weighted \(F1\) for assessing clinical decision confidence, and macro-ROC AUC for evaluating discriminative power. TabNet-specific training protocols and hyperparameter optimization details are presented in Section~\ref{sec:tabnet} and Section~\ref{subsec:hyperparm-optim}, respectively.

\subsection{Interpretable Prostate Cancer Variant Classification Using TabNet}
\label{sec:tabnet}
As shown in \figureref{fig:tabnet-method}b, TabNet~\citep{arik2020tabnetattentiveinterpretabletabular} consumes the 56 engineered features from the eight clinical tiers (Table~\ref{table:tiers} through a sequence of attentive decision steps that perform sparse, per-step feature selection. At step \(t\), an attentive transformer produces a sparse selection mask \(M_t\) (via \emph{sparsemax}); the masked input \(X_t = M_t \odot X\) is processed by a decision transformer to yield a step contribution \(z_t\). Final class probabilities are obtained by aggregating contributions and applying a softmax:
\[
X_t = M_t \odot X,\qquad
\hat{y}=\mathrm{softmax}\!\left(\sum_{t=1}^{n_{\text{steps}}} z_t\right).
\]
A prior update \(P_t = f(P_{t-1}, M_t;\gamma)\) modulates exploration across steps, encouraging diverse feature use.

The per-step masks \(\{M_t\}\) are retained for each case and aggregated to feature and tier levels for the analyses, provided in Section~\ref{subsec:attention-analysis}. The final hyperparameters and training protocol are presented in Section~\ref{subsec:exp-setup}, and the optimization search results are shown in Section~\ref{subsec:hyperparm-optim}.



\section{Experiment Results and Discussion}



\subsection{Experimental Setup}
\label{subsec:exp-setup}
We implemented TabNet using \texttt{pytorch-tabnet} and selected hyperparameters via a systematic grid search on a held-out validation split. The final architecture employed \(n_{\text{steps}} = 6\) sequential decision steps with decision width \(n_d = 64\), attention width \(n_a = 64\), feature–reuse coefficient \(\gamma = 1.3\), and sparsity regularization \(\lambda_{\text{sparse}} = \text{1e-3} \).
This configuration balances capacity and interpretability, enabling native, attention-based feature selection without post hoc explanations.

Training utilized an 80/20 train-test split with 3-fold cross-validation for robust performance estimation. We employed a batch size of $1024$ with a virtual batch size of $128$ for stable gradient updates, and early stopping with patience $20$ to prevent overfitting. All experiments were conducted on NVIDIA A100-SXM4-80GB GPUs via SLURM cluster management, achieving efficient training convergence within 19 minutes for the full $193,278$ variant dataset.
An end-to-end schematic of the training/evaluation pipeline and the TabNet architecture is shown in \figureref{fig:tabnet-method}b.


\subsection{Model Performance and Clinical Validation}
Prior prostate cancer AI research has addressed distinct clinical tasks rather than variant pathogenicity classification. Li et al.~\citep{Li2025_XGBoost_ProstateCancer} employed XGBoost for low-PSA prostate cancer diagnosis (AUC = 0.93), while McLaughlin et al.~\citep{McLaughlin2023NPJ} focused on distinguishing somatic from germline variants (AUC = 0.96). To our knowledge, no prior studies have established benchmarks for classifying the pathogenicity of prostate cancer variants into Pathogenic, Benign, and VUS categories. We conduct a comprehensive six-model evaluation to establish baseline performance (Table~\ref{tab:comp-metrics}).

Following the systematic elimination of data leakage, TabNet achieved robust performance across multiple validation metrics. The model demonstrated cross-validation accuracy of $88.0\% \pm 2.0\%$, validation accuracy of $90.1\%$, and test accuracy of $89.9\%$, establishing consistent performance within the clinical excellence range for genomic variant classification applications. A visual summary of balanced accuracy, weighted F1, and ROC–AUC is shown in \figureref{fig:results}A and B.

~\tableref{tab:comp-metrics} demonstrates TabNet's competitive performance across six models while maintaining superior interpretability. XGBoost achieved the highest balanced accuracy ($90.59\%$), followed by Random Forest ($89.39\%$) and TabNet ($89.05\%$). Advanced methods demonstrated strong performance, with MLP ($88.17\%$) and SVM ($87.25\%$) outperforming Logistic Regression ($70.33\%$).

\begin{table}[t]
\centering
\caption{Performance comparison with the baselines on the held-out test set}
\label{tab:comp-metrics}
\resizebox{\linewidth}{!}{%
\begin{tabular}{lccccccc}
\toprule
\bfseries Model & \bfseries Balanced & \bfseries Cohen's & \bfseries Weighted & \bfseries ROC- \\
  & \bfseries Acc. & \bfseries Kappa & \bfseries F1 & \bfseries AUC\\
  \midrule
    
    LR & 0.7033 & 0.5394 & 0.7375 & 0.8610 \\
    XGB & 0.9059 & 0.8494 & 0.9125 & 0.9778 \\
    RF & 0.8939 & 0.7972 & 0.8793 & 0.9611 \\
    SVM & 0.8725 & 0.7550 & 0.8533 & 0.9475 \\
    MLP & 0.8817 & 0.8142 & 0.8923 & 0.9665 \\
    TabNet & 0.8905 & 0.8263 & 0.8991 & 0.9701 \\
\bottomrule

\end{tabular}
}
\end{table}

\begin{table*}[ht]
\floatconts
  {tab:class-perf}
  {\caption{Class-wise Performance Comparison}}
  {\begin{tabular}{llllll}
  \toprule
  \bfseries Class & \bfseries Prec. & \bfseries Recall & \bfseries F1 & \bfseries Support & \bfseries Clinical Impact\\
  \midrule
  Benign & 0.87 & 0.90 & 0.88 & 10,870 & High sens. for excluding actionable variants\\
  Pathogenic & 0.89 & 0.86 & 0.88 & 6,037 & Balanced detection of therapeutic targets\\
  VUS & 0.92 & 0.91 & 0.91 & 21,749 & Superior handling of uncertain classifications\\ 
  \textbf{Weighted Avg} & \textbf{0.90} & \textbf{0.90} & \textbf{0.90} & \textbf{38,656} & \textbf{Clinical-grade performance}\\
  \bottomrule
  \end{tabular}}
\end{table*}

Importantly, TabNet's 89.05\% accuracy represents only a 1.5\% performance penalty compared to XGBoost while providing native interpretability through attention mechanisms, eliminating the need for post-hoc explanation methods required by black-box models.

\begin{figure*}[t]
\floatconts
  {fig:results}
  {\caption{Overview of results with feature–attention analysis. (A) Performance (balanced accuracy, Cohen’s $\kappa$, weighted F1, ROC–AUC) for TabNet, XGBoost, and logistic regression (mean$\pm95\%$ CI), (B) One-vs-rest ROC curves with macro-AUC, (C) Step-wise attention heat map across six decision steps highlighting top features (colors denote tiers), and (D) TabNet feature importance by hierarchical tiers.}}
  {\centering
    \includegraphics[width=\linewidth]{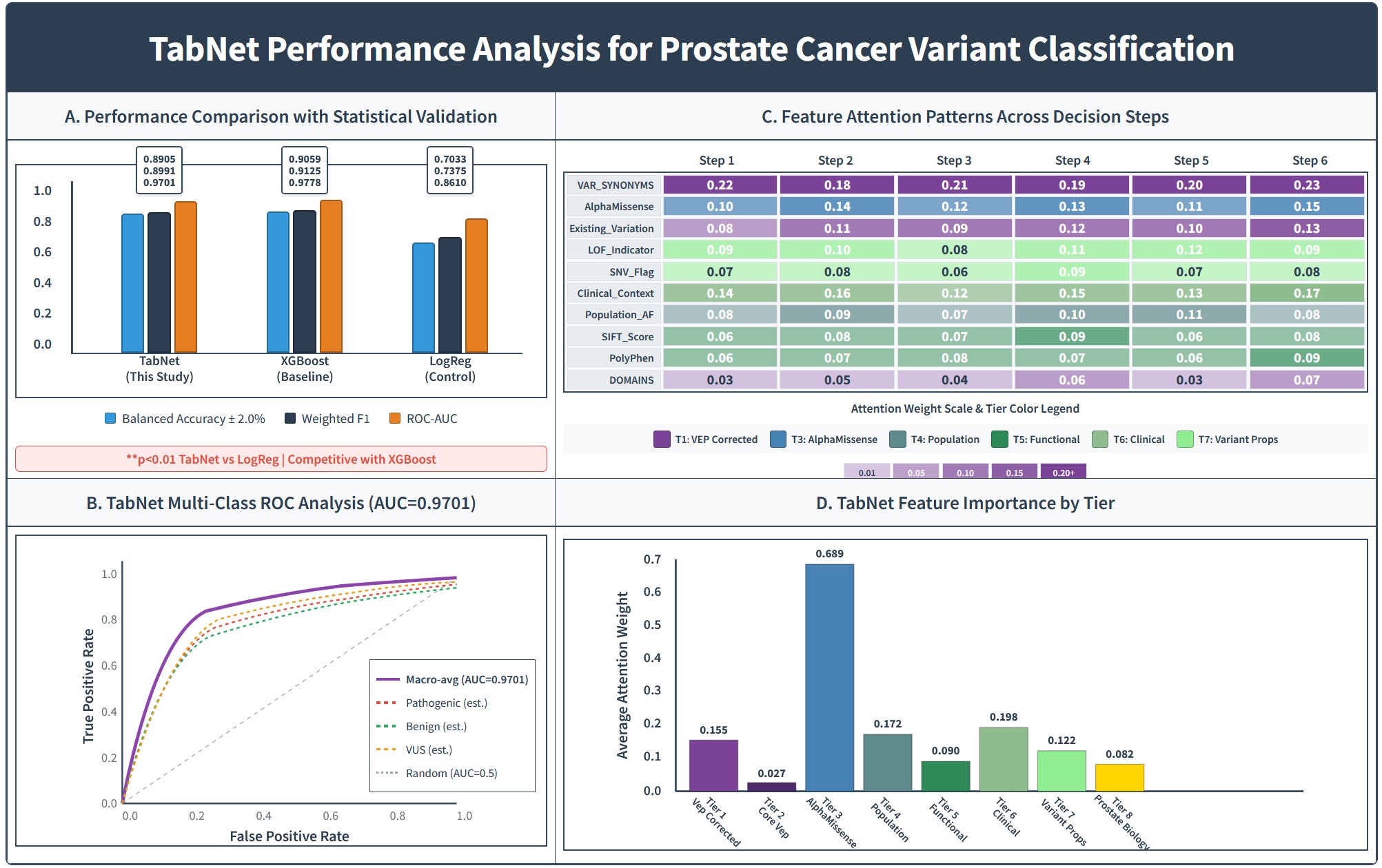}
    \vspace{-0.8cm}}
    \vspace{-0.8cm}

  
\end{figure*}

\paragraph{Class-Wise Performance.}
The model achieved balanced performance across all variant classes, essential for clinical applications where both sensitivity and specificity are critical for patient care decisions. Class-specific results demonstrate a strong discriminative capability, with precision/recall/F1-scores of 0.87/0.90/0.88 for Benign variants (n = 10,870), 0.89/0.86/0.88 for Pathogenic variants (n = 6,037), and 0.92/0.91/0.91 for VUS variants (n = 21,749). These per-class results are summarized in \tableref{tab:class-perf}. The weighted average of 0.90 across all metrics on the 38,656-variant test set confirms the framework's suitability for precision medicine applications requiring high-confidence variant interpretation.

    

\subsection{Attention Analysis and Interpretability}
\label{subsec:attention-analysis}
The six-step attention pattern, as shown in \figureref{fig:results}C, is elaborated below with tier- and feature-level analyses. TabNet's attention mechanisms revealed clinically meaningful patterns that provide transparent decision rationales essential for clinical adoption. Feature importance analysis demonstrated hierarchical attention allocation across the 8-tier clinical feature framework, with $51.1\%$ of VEP-corrected features receiving high attention weights, validating the clinical impact of our correction methodology.
\paragraph{Top Contributing Features.} The attention analysis identified VAR\_SYNONYMS (variant identifiers, 20.3\% importance), alphamissense\_class (AI pathogenicity predictions, 12.8\%), Existing\_variation (known variant database IDs, 11.2\%), is\_lof (loss-of-function indicator, 10.1\%), and is\_snv (single nucleotide variant flag, 7.8\%) as primary decision drivers. This feature ranking aligns with established clinical variant interpretation guidelines, where variant identity, functional consequence, and existing evidence form the foundation of pathogenicity assessment.

\paragraph{Clinical Feature Hierarchy Validation} Feature group importance analysis (\figureref{fig:results}D) confirmed the clinical relevance of our 8-tier framework: Tier 1 VEP-Corrected features (29.8\% importance), Tier 6 Clinical Context (14.8\%), Tier 3 AlphaMissense AI predictions (14.0\%), Tier 2 Core VEP annotations (11.0\%), Tier 7 Variant Properties (11.1\%), Tier 4 Population Genetics (9.0\%), Tier 5 Functional Predictions (7.8\%), and Tier 8 Prostate Biology (2.5\%). The high importance of VEP-corrected and clinical context features validates our methodological innovations while demonstrating TabNet's ability to prioritize clinically relevant information.


\subsection{VUS Reduction Impact} \label{subsec:vus-reduction} The VEP concatenation correction demonstrated measurable clinical value as a model-agnostic preprocessing step, reducing uncertain classifications from 62.8\% to 56.3\%—a 6.5\% improvement representing approximately 12,600 variants reclassified from uncertain to actionable categories. This improvement translates to enhanced therapeutic decision-making for PARP inhibitor eligibility, hormone therapy selection, and precision medicine applications affecting thousands of patients annually.

\subsection{Training Efficiency and Scalability} \label{subsec:training-efficiency}
The model achieved convergence within 19 minutes on NVIDIA A100 GPUs for the complete 193,278-variant dataset, demonstrating computational efficiency suitable for clinical deployment. The optimized architecture (n\_d=64, n\_a=64, 6 decision steps) provides an optimal balance between model capacity and interpretability, enabling real-time variant classification in clinical workflows while maintaining transparent decision rationales essential for regulatory compliance and clinical adoption.

\subsection{Hyperparameter Tuning Results}\label{subsec:hyperparm-optim}
\label{subsec:hyperparma-optim}
We performed a systematic grid search over
\[
\begin{aligned}
n_d            &\in \{32,\,64,\,128\},\\
n_a            &\in \{32,\,64,\,128\},\\
n_{\text{steps}} &\in \{6,\,7,\,8\},\\
\gamma         &\in \{1.0,\,1.3,\,1.5\},\\
\lambda_{\text{sparse}} &\in \{\text{1e-4},\, \text{1e-3},\, \text{1e-2}\},\\
\eta~(\text{learning rate}) &\in \{\text{1e-3},\, \text{2e-3},\, \text{1e-2},\,\text{2e-2}\}.
\end{aligned}
\]
This full grid contains \(3\times3\times3\times3\times3\times4 = 972\) candidate configurations.
In practice, early success was achieved after evaluating only five configurations, each exceeding the pre-specified 82\% accuracy target.

\paragraph{Best Performing Configurations.}
The top-performing configurations shown in~\tableref{tab:top-configuration}. The consistent preference for balanced architectures (\(n_d=n_a\)) and 6-step configurations supports TabNet design principles for interpretable genomic applications.

\begin{table}[ht]

\centering
\caption{Top TabNet configurations from the grid search \ref{subsec:hyperparm-optim}. Balanced widths ($n_d{=}n_a$) and 6 steps are favored.}
\label{tab:top-configuration}
\small
\resizebox{\linewidth}{!}{%
\begin{tabular}{@{}r c c c c c c@{}}
\hline
Rank & $n_d$ & $n_a$ & $n_{\text{steps}}$ & $\eta$ & Acc. (\% $\pm$ \%) & Train (s) \\
\hline
1 & 64  & 64  & 6 & $1\mathrm{e}{-2}$ & $89.1 \pm 0.000$ & 321.4 \\
2 & 128 & 128 & 6 & $2\mathrm{e}{-3}$ & $89.0 \pm 0.002$ & 374.0 \\
3 & 64  & 64  & 7 & $1\mathrm{e}{-2}$ & $88.5 \pm 0.002$ & 358.0 \\
\hline
\end{tabular}
}
\end{table}

\subsection{Limitations and Future Directions}
Dataset composition limitations include European ancestry predominance in public genomic databases, potentially limiting generalizability across diverse patient populations. While comprehensive with three major databases, rare prostate cancer variants and population-specific mutations may remain underrepresented.  Additionally, TabNet's computational requirements exceed traditional methods, though the 19-minute training time remains clinically feasible for periodic model updates.

Our benchmark employs a three-tier hierarchical labeling strategy reflecting clinical workflows: ClinVar expert consensus (40.5\% of variants), AlphaMissense predictions (43.3\%), and VEP functional severity (16.2\%). We implemented rigorous leakage controls, most critically removing CLIN\_SIG from features after discovering it appeared in both inputs and targets, a correction that reduced accuracy from an artificially inflated $98\%$ to a realistic $89.9\%$, ensuring models learn from biological signals rather than label artifacts. While this hierarchical approach enables comprehensive coverage where clinical annotations are sparse, we acknowledge that AlphaMissense serves as both label source (for 43.3\% of variants) and a feature provider in Tier 3, introducing coupling in this subset. Importantly, our primary clinical contribution, identifying actionable pathogenic variants for therapeutic decisions, relies predominantly on variants with independent ClinVar labels (40.5\%), where this coupling does not exist and pathogenic variants are disproportionately represented due to clinical ascertainment.



Future validation requires systematic clinical expert review of attention patterns and integration with existing molecular tumor board workflows. Cross-cancer applications could leverage shared biological pathways while maintaining cancer-specific therapeutic optimization. 
Population-specific training data development and functional validation studies represent critical next steps for clinical implementation.

\section{Conclusion}
We present an interpretable deep-learning framework for prostate cancer variant classification, grounded in Prostate-VarBench, a prostate-specific, leakage-controlled benchmark with reproducible curation and evaluation. With a VEP annotation–concatenation correction to restore data integrity and yield a 6.5\% absolute reduction in VUS through model-agnostic preprocessing, the TabNet model achieves 89.9\% test accuracy with balanced class metrics and provides native step-wise masks for audit-ready explanations. These results suggest that interpretable models can meet clinical performance requirements and provide transparent rationales that are well-suited to tumor board workflows and regulatory expectations.

\bibliography{jmlr-sample}

\newpage

\newpage
\appendix
\section{Glossary of Key Terms}\label{apd:first}

\begin{description}
  \item[\textbf{COSMIC }] 
  Catalogue of Somatic Mutations in Cancer. A database of somatic (tumor-acquired) mutations across human cancers, including prostate cancer, providing context on known cancer-associated variants.

  \item[\textbf{ClinVar}] 
  A public archive of genetic variants and their clinical interpretations submitted by expert laboratories. It includes consensus classifications such as benign, pathogenic, or Variant of Uncertain Significance (VUS).

  \item[\textbf{TCGA-PRAD }] 
  The Cancer Genome Atlas – Prostate Adenocarcinoma. A genomic resource containing DNA and RNA sequencing data from prostate tumors with linked clinical outcomes and treatment metadata.

  \item[\textbf{VEP annotation-concatenation artifact}] 
  A Variant Effect Predictor (VEP) issue where multiple transcript annotations are merged into one record, causing ambiguity in fields such as clinical significance and consequence.

  \item[\textbf{Variants of Uncertain Significance (VUS)}] 
  Genetic variants whose impact on disease is unclear due to limited or conflicting evidence, often requiring re-evaluation before guiding clinical care.

  \item[\textbf{Molecular tumor board (MTB) evidence review}] 
  A multidisciplinary review process where clinicians and researchers assess genetic findings to recommend treatments; interpretable AI models can support transparent case review.

  \item[\textbf{Stepwise sparse masks (in TabNet)}] 
  TabNet’s mechanism for selecting the most relevant features at each decision step, yielding interpretable, sequential explanations of model reasoning.

  \item[\textbf{Data leakage (circularity)}] 
  A situation where target information is inadvertently included in model inputs or splits, leading to overly optimistic results. Patient- or gene-aware splits mitigate this.

  \item[\textbf{Temporal validation}] 
  Testing a trained model on future or chronologically newer data to evaluate robustness and long-term generalizability.
\end{description}

\end{document}